\documentclass{emulateapj}
\lefthead{Dewangan \& Griffiths}
\slugcomment{To appear in ApJ {\it Letters}}

\newcommand\asca{{\it ASCA}}
\newcommand\sax{{\it BeppoSAX}}
\newcommand\chandra{{\it Chandra}}

\newcommand\xmm{{\it XMM-Newton}}

\newcommand\s{{\rm~s}}
\newcommand\kev{{\rm~keV}}
\newcommand\ev{{\rm~eV}}
\newcommand\kms{\ifmmode {\rm~km\ s}^{-1} \else ~km s$^{-1}$\fi}
\newcommand\ks{\ifmmode {\rm~ks} \else ~ks\fi}
\newcommand\Hunit{\ifmmode {\rm~km\ s}^{-1}\ {\rm Mpc}^{-1}
        \else ~km s$^{-1}$ Mpc$^{-1}$\fi}
\newcommand\ctssec{\ifmmode {\rm~count\ s}^{-1} \else ~count s$^{-1}$\fi}
\newcommand\ergsec{\ifmmode {\rm~erg\ s}^{-1} \else
        ~erg s$^{-1}$\fi}
\newcommand\funit{\ifmmode {\rm~erg\ s}^{-1}\;{\rm cm}^{-2} \else
        ~ergs s$^{-1}$ cm$^{-2}$\fi}
\newcommand\phflux{\ifmmode {\rm~photon\ s}^{-1}\;{\rm cm}^{-2}
        \else   ~photon s$^{-1}$ cm$^{-2}$\fi}
\newcommand\efluxA{\ifmmode {\rm~erg\ s}^{-1}\;{\rm cm}^{-2}\;{\rm
        \AA}^{-1} \else ~erg s$^{-1}$ cm$^{-2}$ \AA$^{-1}$\fi}
\newcommand\efluxHz{\ifmmode {\rm~erg\ s}^{-1}\;{\rm cm}^{-2}\;{\rm
        Hz}^{-1} \else ~erg s$^{-1}$ cm$^{-2}$ Hz$^{-1}$\fi}
\newcommand\cc{\ifmmode {\rm~cm}^{-3} \else cm$^{-3}$\fi}
\newcommand\FWHM{\ifmmode {\rm~FWHM} \else ${\rm~FWHM}$\fi}
\newcommand\Msun{\ifmmode M_{\odot} \else $M_{\odot}$\fi}
\newcommand\Lsun{\ifmmode L_{\odot} \else $L_{\odot}$\fi}

\newcommand\hbeta{\ifmmode {\rm H}\beta \else H$\beta$\fi}
\newcommand\Kalpha{\ifmmode {\rm K}\alpha \else K$\alpha$\fi}
\newcommand\NH{\ifmmode N_{\rm H} \else N$_{\rm H}$\fi}
\usepackage{graphicx}
\usepackage{here}

\begin{document}

\title{Type 2 counterparts of narrow-line Seyfert~1 galaxies}
\author{G. C. Dewangan \& R. E. Griffiths} \affil{Department of
  Physics, Carnegie Mellon University, 5000 Forbes Ave, Pittsburgh, PA
  15213 USA} \email{gulabd@cmu.edu}
\begin{abstract}
  Unified models of Seyfert galaxies, based on viewing angles,
  successfully explain the observed differences between type 1 and 2
  Seyferts. The existence of a range in accretion rates ($\dot{m} \sim
  0.001-1$) relative to the Eddington rate (from broad-line Seyfert~1s
  to narrow-line Seyfert~1s or NLS1s) and the unification of Seyfert
  galaxies imply that there must be type 2 counterparts of NLS1
  galaxies i.e., Seyfert 2s with high accretion rate or small black
  hole mass. One such Seyfert 2, NGC~5506, has already been unmasked
  based on near infra-red spectroscopy. Here we confirm the above
  result, and present evidence for two additional type 2 counterparts
  of NLS1s based on \xmm{} observations. The three AGNs NGC~7314,
  NGC~7582 and NGC~5506, with a type 1.9/2 optical spectrum, show
  extremely rapid variability by factors $>2.4$, $\sim 1.3$, and $\sim
  1.7$ in $200\s$, $350\s$ and $300\s$, respectively, and steep
  $2-12\kev$ spectrum ($\Gamma \ga 2$) in their intrinsic X-ray
  emission, characteristic of NLS1 galaxies. These observations
  establish the {\it `obscured NLS1 galaxies'} as a subclass of
  Seyfert 2 galaxies.
\end{abstract}

\keywords{accretion, accretion disks -- galaxies: active -- X-rays:
  galaxies }

\section{Introduction}
Seyfert 1 galaxies exhibit a large range in their observed
characteristics e.g., the FWHM of the H$\beta$ line in the range of
$\sim1000-10000{\rm~km~s^{-1}}$, X-ray variability timescale from a
few hundred seconds to years, the soft ($0.1-2\kev$) and hard
($2-10\kev$) X-ray photon indices in the range of $\sim 1.0 - 4.0$
and $\sim 1.5 - 2.5$, respectively, etc.  The Seyfert 1s  with
extreme properties, i.e.  $FWHM_{H\beta} \la 2000\kms$, steepest
X-ray spectra and rapid X-ray variability form
a distinct subclass known as narrow-line Seyfert 1 galaxies (NLS1s).
The standard unification scheme of Seyfert galaxies supposes that the
central engines of Seyfert 1s and 2s are basically the same, and the
differences in their observed properties are due to different viewing
angles. In Seyfert 2s, our line of sight passes
through a cold torus which obscures the broad line region and the
accretion disk (Antonucci 1993).  Now it has been established that
Seyfert 1 galaxies span a large range $\sim 0.001 - 1$ in their
relative accretion rates, NLS1s lying at the higher end of the
relative accretion rate distribution (Pounds et al. 1995). These facts
imply that there must be Seyfert 2s with high relative accretion rates
similar to those of NLS1s i.e., the type 2 counterpart of NLS1s. These
objects are largely missing at present due to the difficulty in
isolating them using the conventional observational techniques of the
optical regime.  The only way to distinguish them is to use
observations  which are not affected by absorption such as
near infra-red spectroscopy, optical spectropolarimetry and X-ray
emission above 2 keV.

In this {\it Letter}, we present X-ray evidence for three type 2
counterparts (NGC~7582, NGC~5506 and NGC~7314) of NLS1s.
NGC~7582 is a classical Seyfert 2, with well defined [O~III]
cone (Sosa-Brito et al. 2001; Aretxaga et al. 1999; Storchi-Bergmann
\& Bonatto 1991). Aretxaga et al. (1999) observed NGC~7582 in an
unusual optical state during July 1998 when it showed broad components
of H$\alpha$, H$\beta$ ($FWHM \sim 12000\kms$) blue-shifted by $\sim
2500\kms$.  The flux state was relatively high during the optical
event, and then decreased gradually to its typical type 2 state in
October 1998. Aretxaga et al. (1999) examined three scenarios: capture
of a star by a super-massive black hole, a reddening change in the
surrounding torus, and the radiative onset of a Type IIn supernova in
the nuclear regions. These authors raise serious concerns with the
first two models, and favor the SN theory to explain the broad lines.
\sax{} study showed that X-ray emission from NGC~7582 is consistent
with the Seyfert 2 nature, and a single  component dominates
the $2-100\kev$ band (Turner et al. 2000).  Changes in nuclear X-ray
flux appeared to be uncorrelated to the gradual decline in the optical
flux noted after the high state of July 1998.  NGC~5506 ($z=0.0065$)
has been classified as a Seyfert 2 (de Robertis \& Osterbrock 1986).
Shuder (1980) reported to have found a weak, broad component to
H$\alpha$. However, both Veron et al. (1980) and Whittle (1985) found
no such component. Nagar et al. ( 2002) discovered a broad
(FWHMM$<2000{\rm~km~s^{-1}}$) O~{I}$\lambda 1.1287{\rm \mu m}$ line
and the $1{\rm~\mu m}$ Fe II lines in the near-IR spectrum of
NGC~5506, thus unmasking an `obscured NLS1 nucleus' in a Seyfert 2.  Bianchi et al. (2003) presented the recent X-ray history of
NGC~5506, the overall picture consists of a nucleus absorbed by cold
gas with $N_H \simeq 10^{22}{\rm~cm^{-2}}$ and surrounded by a
Compton-thick torus.  NGC~7314 has an optical spectrum and line fluxes
typical of a Seyfert 2  (Morris \& Ward 1988; Winkler 1992;
Joguet et al. 2001).  An {\it HST} spectrum of the nuclear region of
NGC~7314 revealed  broad component of H$\alpha$, consistent with a
Seyfert 1.9 classification (Hughes et al. 2003). NGC~7314 is well
known for its rapid, large amplitude X-ray variability (Turner et al.
1987; Yaqoob et al. 1996). Yaqoob et al. (2003) detected variable,
redshifted narrow lines from He-like and H-like iron using \chandra{}
HETG spectrum of NGC~7314.

 \section{Observations \& Data Reduction}
 The three AGNs were observed with \xmm{}. NGC~7582 was observed for
 $\sim 20\ks$ on 25 May 2001 using the medium filter and in the full
 frame mode of the EPIC instruments. NGC~5506 has been observed five
 times between 2001 and 2004. Here we used the first observation
 carried out on 3 February 2001 for $20\ks$ using the medium filter
 and large window mode of EPIC instruments.  NGC~7314 was observed on
 2 May 2001 twice for $10\ks$ and $44\ks$, separated by $10{\rm~hr}$
 using the medium filter and in the full frame mode for MOS and small
 window model for the PN. Here we use the longer observation.  The
 data were processed and filtered using {\tt SAS v6.1.0}. None
 of the three observations is severely affected with high particle
 background. Source spectra and light curves were
 extracted from circular regions of radii 40$\arcsec$ centered on the
 source positions, with event patterns 0--4 for the PN and 0--12 for
 the MOS. Background PN spectra and light curves for NGC~7582 and
 NGC~5506 were extracted from nearby, source free, regions located at
 approximately the same readout position as that for the source
 circle. In the case of NGC~7314, we chose the background regions in
 the available source free region due to the small window mode of the
 PN observation. We used nearby source free circular background
 regions for the MOS.

\begin{figure}
  \centering \includegraphics[width=8.5cm]{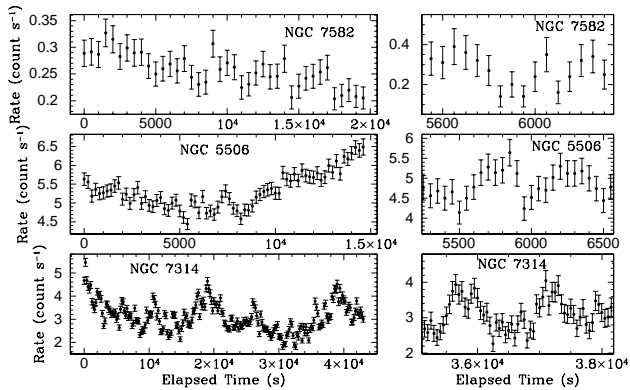}
  \caption{The $2-12\kev$ EPIC PN light curves of NGC~7582, NGC~5506,
    and NGC~7314. The left panels show light curves spanning the full
    lengths of the observations with bins of $500\s$ (NGC~7314) and
    $200\s$ (NGC~5506 \& NGC~7314). The right panels show the extreme
    variability events with smaller bins of $50\s$.}
  \label{f1}
\end{figure}

\section{X-ray Variability}
Figure~\ref{f1} shows the background corrected $2-12\kev$ PN light
curves of NGC~7582, NGC~5506 and NGC~7314.
The left panels show the light curves spanning the full length of the
observations.
The right panels show the extreme variable events with
smaller bins. 
NGC~7582 showed a gradual decline in its X-ray
intensity from a count rate of $\sim 0.3$ to $\sim
0.2{\rm~counts~s^{-1}}$ in addition to small amplitude rapid
variability. The most extreme variability occurred when the count rate
decreased by $0.19\pm0.056{\rm~counts~s^{-1}}$ from an elapsed time of
$5700\pm50\s$ to $5900\pm50\s$. We used an average of three data
points in all calculations of count rate variability. Other extreme
variable events include a decrease in count rate of
$0.16\pm0.04{\rm~counts~s^{-1}}$ within $600\s$ during the elapsed
time interval of $14300-14900\s$, and an increase by
$0.10\pm0.038{\rm~counts~s^{-1}}$ within $500\s$ during the interval
$14900-15400\s$. 
 NGC~5506 shows flickering behavior during the first half
and then gradually brightens in the second half of the observation.
The most extreme variable events include an increase in the count rate
by $0.89\pm0.25$ within $350\s$, followed by a rapid decline by a
similar amplitude within just $150\s$ and then again an increase in
the count rate similar to the initial event.  Among the three AGNs,
NGC~7314 shows the most remarkable rapid variability throughout the
observation. The most extreme event occurred around $35500\s$ after
the start of the observation, the count rate increased by
$1.25\pm20{\rm~counts~s^{-1}}$ within just $300\s$.

The remarkably rapid variability mentioned above are even more extreme
intrinsically. The $2-12\kev$ X-ray emission from Seyfert 1.9/2
galaxies consists mainly of two components: ($i$) absorbed primary
continuum, and ($ii$) Compton reflection from distant cold material.
The reflection component is relatively steady over timescales of years. The
$2-10\kev$ emission from Compton-thick Seyfert 2s is almost
entirely Compton reflection emission, while both primary  and
Compton reflection emission contribute to the X-ray emission from Compton-thin
Seyfert 2s. Below we estimate these contributions 
 with spectral modeling.

\section{X-ray Spectra}
In the following analysis we consider the time-averaged EPIC
PN and MOS spectra of NGC~7582 and PN spectra of NGC~7314. We did not
use the MOS data for NGC~7314 due to the high signal-to-noise of the
PN data and slight mismatch in the continuum shapes inferred from the
PN and MOS data. The spectra were binned to a minimum of 20 counts per
channel and analyzed with {\tt XSPEC 11.3}. 
Only the data above 2 keV are considered here. A detailed study of the
complex soft X-ray emission is beyond the scope of this {\it Letter},
and has no significant impact on the final results.

\begin{figure}
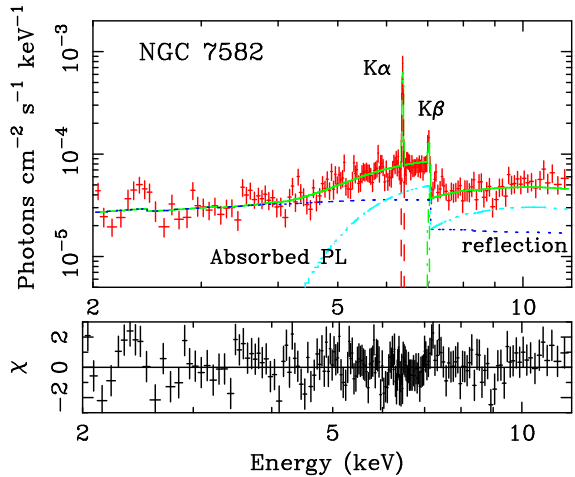

  \centering
  \includegraphics[angle=-90,width=7.5cm]{f2a.ps}
  \includegraphics[angle=-90,width=7.5cm]{f2b.ps}
  \caption{{\it Top:} Unfolded EPIC PN X-ray spectrum of NGC~7582 and
    best-fit spectral model consisting of an absorbed power-law,
    Compton reflection and iron K lines. {\it Bottom:} Deviations, in
    units of sigma, of the observed data from the best-fit model.}
  \label{f2}
\end{figure}

\begin{figure}
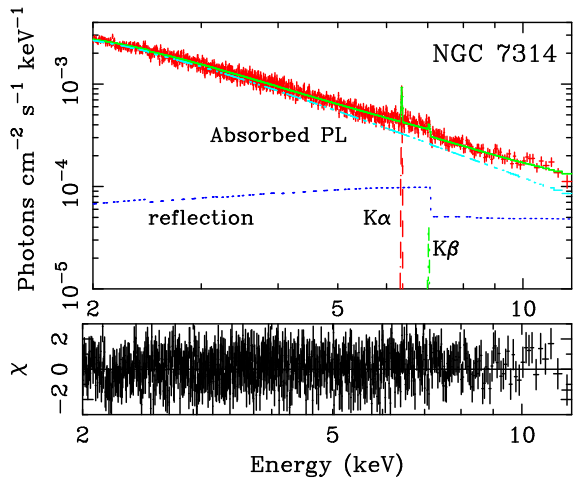

  \centering
  \includegraphics[angle=-90,width=7.5cm]{f3a.ps}
  \includegraphics[angle=-90,width=7.5cm]{f3b.ps}
  \caption{Same as Fig.~\ref{f2} but for NGC~7314.}
  \label{f3}
\end{figure}

The X-ray spectra of a Seyfert 2s generally show signatures of
reprocessed emission; namely a strong iron K$\alpha$ line at $\sim
6.4\kev$ and a corresponding absorption edge at $\sim 7.1\kev$. In
view of these features, we adopt a baseline model consisting of a
heavily absorbed power-law, a Compton reflection component
from neutral material and a narrow Gaussian to represent the iron
K$\alpha$ line. For the sake of completeness, we also added a second
Gaussian line to describe the iron K$\beta$ line from neutral iron.
The energy and strength of the K$\beta$ line were fixed at their
theoretical values of $7.06\kev$ and $10\%$ of the strength of the
corresponding K$\alpha$ line, respectively. All the components were
modified by the Galactic column, listed in Table~\ref{t1}. We used the
reflection model {\it pexrav} (Magdziarz \& Zdziarski 1995). This
model calculates the reflected spectrum from a neutral disk exposed to
an exponentially cutoff power-law continuum.  The {\it pexrav} model
was set to produce the reflected emission only, while the power-law
was set to provide the primary continuum. We fix the cutoff energy of
the primary power law at $100\kev$, disk inclination at $60\deg$, and
the abundance of heavy elements at the solar value.  The free
parameters are the photon index and the normalization of the primary
power law, and the relative amount of reflection compared with the
directly viewed primary spectrum ($R$). We also varied the intrinsic
absorption column ($N_H^{int}$), the line energy ($E_{FeK\alpha}$) and
the flux ($f_{FeK\alpha}$) of the iron K$\alpha$ line. The baseline
model resulted in a good fit ($\chi^2 = 361.2$ for $330$ degrees of
freedom (dof) for NGC~7582 and $\chi^2 = 1335.0$ for $1424$ dof for
NGC~7314). The best-fit parameters are listed in Table~\ref{t1}, and
the unfolded spectra and deviations are shown in Figures~\ref{f2}
(NGC~7582) and \ref{f3} (NGC~7314). The contribution of the Compton
reflection component, including the narrow iron lines, to the observed
X-ray flux in the $2-12\kev$ band is $48.8\%$ for NGC~7582 and
$14.1\%$ for NGC~7314.


A comparison of \asca{} (Xue et al. 1998), \sax{} (Turner et al. 2000)
and \xmm{} observations of NGC~7582 reveals a continuous increase in
the absorption column ($\Delta N_H \sim 8\times10^{23}{\rm~cm^{-2}}$)
and the $2-10\kev$ photon index ($\Delta \Gamma \sim 0.9$) from 1994
to 2001.  NGC~7582 was in the lowest flux state during the \xmm{}
observation in May 2001.  The observed $2-10\kev$ flux ($\sim 4\times
10^{-12}{\rm~erg~cm^{-2}~s^{-1}}$) is a factor $\sim 5$ lower than that
observed with the \sax{} in the same band. We cannot confirm the broad
asymmetric iron K$\alpha$ line from NGC~7314 detected with \asca{}
(Yaqoob et al. 1996). \xmm{} is not suitable to detect the multiple,
redshifted narrow iron lines present in the \chandra{} HETG spectrum
(Yaqoob et al. 2003). The best-fit photon index ($\Gamma \sim 2.2$) to
the \xmm{} data is similar to that derived from the \sax{}
data ($\Gamma \sim 2.1$; Risaliti 2002).

\begin{table}
  \caption{Best-fit parameters \label{t1}}
  \begin{tabular}{lll}
    \hline \hline
    Parameter                      & NGC~7582 & NGC~7314 \\ \hline
    $N_H^{Gal}$ ($10^{20}{\rm~cm^{-2}}$)  & $1.93$(f)              & $1.46$(f)     \\
    $R$                            & $1.84_{-0.34}^{+0.46}$   & $2.83_{-0.44}^{+0.53}$         \\
    $i$                            &  $60\deg$(f)            &   $60\deg$(f)  \\
    $N_H^{int}$ ($10^{22}{\rm~cm^{-2}}$)   & $89.6_{-6.3}^{+8.7}$    & $1.05_{-0.14}^{+0.13}$ \\
    $\Gamma$                       & $2.26_{-0.17}^{+0.14}$   & $2.19_{-0.06}^{+0.09}$  \\
    $E_{FeK\alpha}$($\kev$)          & $6.41_{-0.01}^{+0.01}$   &  $6.38_{-0.04}^{+0.04}$     \\
    $\sigma$($\ev$)                 & $10$(f)               & $10$(f)    \\
    $f_{FeK\alpha}$ ($10^{-5}{\rm photons~cm^{-2}~s^{-1}}$) & $1.87_{-0.2}^{+0.2}$   & $1.39_{0.48}^{+0.34}$    \\
    $EW$($\ev$)                     & $521_{-141}^{+139}$      &  $147_{-109}^{+128}$ \\ 
    $E_{FeK\beta}$ ($\kev$)           & $7.06$                &  $7.06$(f)       \\
    $\sigma$($\ev$)                 & $10$(f)                &  $10$(f)            \\
    $f_{FeK\beta}$ ($10^{-5}{\rm photons~cm^{-2}~s^{-1}}$)  & $0.19$                &  $0.14$  \\
    $EW$($\ev$)                     & $66$                   &   $18$   \\ 
    $f_{refl}$~\tablenotemark{a}    & $2.76$                 & $7.20$   \\ 
    $f_{PL}$~\tablenotemark{a}     &  $2.89$               & $3.69$   \\
    $f_{refl}^{int}$~\tablenotemark{a} & $2.77$               & $7.20$  \\ 
    $f_{PL}^{int}$~\tablenotemark{a}  & $24.4$               & $40.8$  \\  
    $\chi^2/dof$                         & $361.2/330$      &  $1334.9/1424$   \\ \hline
  \end{tabular}
\tablenotetext{a}{Observed or  intrinsic flux in units of $10^{-12}{\rm~erg~cm^{-2}~s^{-1}}$.}
\end{table}

\begin{table}
\caption{$2-12\kev$ Count rate and unabsorbed power-law flux during the most rapid  variable events. \label{t2}}
{\footnotesize
\begin{tabular}{lccc}
\hline \hline
Object & Elapsed time & Count rate                        & unabsorbed PL flux \\
       & ($\s$)       & (${\rm counts~s^{-1}}$)      & (${\rm erg~cm^{-2}~~s^{-1}}$) \\
\hline
NGC~7582 & $5700\pm50$   & $0.35\pm0.05$ & $3.37_{-0.81}^{+0.81}\times10^{-11}$ \\
         & $5900\pm50$   & $0.16\pm0.03$ & $0.28_{-0.28}^{+0.77}\times10^{-11}$ \\
         & $14300\pm50$  & $0.34\pm0.03$ & $3.21_{-0.49}^{+0.49}\times10^{-11}$   \\
         & $14900\pm50$  & $0.18\pm0.02$ & $0.60_{-0.32}^{+0.32}\times10^{-11}$   \\
NGC~5506 & $5500\pm50$   & $4.43\pm0.17$ & $5.22_{-0.26}^{+0.26}\times10^{-11}$   \\
         & $5850\pm50$   & $5.32\pm0.19$ & $6.58_{-0.26}^{+0.26}\times 10^{-11}$    \\
         & $6000\pm50$   & $4.50\pm0.17$ & $5.34_{-0.25}^{+0.25}\times10^{-11}$  \\
NGC~7314 & $35350\pm50$  & $2.53\pm0.13$ & $1.91_{-0.14}^{+0.14}\times10^{-11}$  \\
         & $35650\pm50$  & $3.78\pm0.16$ & $3.24_{-0.13}^{+0.13}\times10^{-11}$   \\  \hline
       \end{tabular}}
   \end{table}


NGC~5506 was observed simultaneously with \xmm{} and \sax{} between
February 1 and 3, 2001. A joint spectral analysis of the two data sets
was performed by Matt et al. (2001). The best-fit model to the PN and
PDS data was similar to the baseline model except for an additional broad
K$\alpha$ line from ionized iron. The primary power-law was steep
($\Gamma = 1.98_{-0.02}^{+0.05}$) and heavily absorbed ($N_H^{int} =
3.44_{-0.12}^{+0.13}\times 10^{22}{\rm~cm^{-2}}$). Using the best-fit
parameters derived by Matt et al. (2001), we estimated that the
contribution of the Compton reflection including the unresolved
K$\alpha$ line from neutral iron is $17.6\%$ to the total observed
X-ray emission ($f_{X} = 6.76\times 10^{-11}{\rm~erg~cm^{-1}~s^{-1}}$)
in the $2-12\kev$ band, while the absorption corrected contribution is
$14.4\%$ to the intrinsic X-ray emission ($f_X^{int} =8.28\times
10^{-11}{\rm~erg~cm^{-2}~s^{-1}}$) in the $2-12\kev$ band.

\section{Intrinsic X-ray Variability}
To estimate the  variability amplitudes of the primary power-law, we have corrected the observed count rates for the
contribution of the Compton reflection and heavy obscuration. First,
we converted the count rates into absorbed
power-law flux using a conversion factor between the time-averaged
count rate and observed flux, and the contribution of the reflection
emission as derived from the spectral modeling. The absorbed power-law
flux were then converted to unabsorbed power-law flux using the
best-fit photon indices and the absorption columns (see
Table~\ref{t1}). The count rates and the unabsorbed power-law flux at
the beginning and the end of the rapid variability events are listed
in Table~\ref{t2}. The $2-12\kev$ unabsorbed flux of the primary power
law varied by a factor $\ga2.4$ or a change in flux, $\Delta f
>1.5\times 10^{-11}{\rm~erg~cm^{-1}}$ in $200\pm71\s$ in the case of
NGC~7582. A change in the intrinsic power-law flux $\Delta f =
(1.36\pm0.37)\times 10^{-11}{\rm~erg~cm^{-2}~s^{-1}}$ in $250\pm71\s$
has been observed from NGC~5506. A similar change, $\Delta f =
(1.33\pm 0.19)\times 10^{-11}{\rm~erg~cm^{-2}~s^{-1}}$ in
$300\pm71\s$, is evident from NGC~7314.

\section{Discussion}
We have studied the temporal and spectral characteristics of three
obscured Seyfert galaxies.
NGC~7582 showed a factor of two variability in the $2-12\kev$ count
rate or a change of $2-12\kev$ intrinsic luminosity, $\Delta L >
9.3\times10^{41}{\rm~erg~s^{-1}}$ of the primary emission in just
$\sim 200\s$. NGC~5506 showed a change in the count rate of
$0.89\pm0.25{\rm~counts~s^{-1}}$ or a change in the luminosity of
$\Delta L = (1.1\pm0.3)\times10^{42}{\rm~erg~s^{-1}}$ in an interval
of $250\s$ only. NGC~7413 showed a change in the count rate of
$1.25\pm0.20{\rm~counts~s^{-1}}$ or a change in the intrinsic primary
luminosity $\Delta L = (6.7\pm0.9)\times10^{41}{\rm~erg~s^{-1}}$ in
the $2-12\kev$ band.  These are the most extreme rapid variability
events ever observed from Seyfert 1.9/2 galaxies.  Application of the
efficiency ($\eta$) limit: $\eta > 4.8\times10^{-43}\Delta L/\Delta t$
(Fabian 1979), however, shows that the these events do not violate the
limit for a Swarzschild black hole.  Among the radio-quiet AGNs, the
most rapid X-ray variability on timescales of a few hundred seconds is
observed from NLS1s (see e.g., Boller et al. 1996;
Leighly 1999a). The rapid X-ray variability observed from NGC~7582,
NGC~5506 and NGC~7314 are similar to that of NLS1s.

   In addition to the extreme variability, the three AGNs show
   steep $2-12\kev$ primary power-law  (NGC~7582: $\Gamma =
   2.26_{-0.17}^{+0.14}$; NGC~5506: $\Gamma = 1.98_{-0.02}^{+0.05}$;
   NGC~7314: $\Gamma = 2.19_{-0.06}^{+0.09}$). These photon indices
   are steeper than the mean index, $\Gamma = 1.79\pm0.01$, of
   the $3-200\kev$ intrinsic power-law for 20 bright Compton-thin
   Seyfert 2s (Risaliti 2002). The photon indices are also
   steeper than the mean index of $1.78\pm0.11$ for the
   broad-line Seyfert 1s but similar to the mean index
   $\Gamma = 2.19\pm0.10$ for NLS1s in the sample of Leighly
   (1999b).

   The similarity of the extreme variability and steep primary
   power-law emission of the three obscured Seyfert galaxies and that
   of NLS1s strongly suggests that the accretion disk-black hole
   system, responsible for the primary X-ray emission, is similar in
   the above two types of Seyfert galaxies. Many of the observed
   characteristics of NLS1s are interrelated e.g., the width
   of the H$\beta$ line is anti-correlated with the soft ($0.1-2\kev$)
   and hard ($2-10\kev$) power-law photon indices (Boller et al. 1996;
   Brandt, Mathur \& Elvis 1997). Thus if the central engines of NLS1s
   and that of NGC~7582, NGC~5506 and NGC~7314 are similar, then
   permitted optical lines from the BLR should be narrower (FWHM$\la
   2000\kms$), similar to that of NLS1s. Indeed, the discovery of
   near-IR permitted  O~ {I}$\lambda 1.1287{\rm \mu m}$ line 
   (FWHM$<2000{\rm~km~s^{-1}}$) and $1{\rm~\mu}$ Fe II lines  by 
\citet{Nagaretal02} is fully
   consistent with the X-ray nature of NGC~5506.  The dramatic
   appearance of broad Balmer lines in the optical spectrum of
   NGC~7582 in July 1998 is not expected from an obscured NLS1.
   However, the broad lines may not be related to the AGN, as the
   evolution of H$\alpha$ line width closely resembled with that of
   SN~1988Z (Aretxaga et al. 1999). A near infra-red spectrum of
   NGC~7582 during its typical type 2 state will further clarify its
   nature.

   In view of the above results, we suggest that the type 1.9/2 AGNs
   NGC~7582, NGC~5506 and NGC~7314 are the type 2 counterpart of NLS1
   galaxies. As NLS1s form a special subclass of Seyfert 1 galaxies
   due to their extreme characteristics, the Seyfert 1.9/2s with
   extreme X-ray characteristics form a special subclass of Seyfert
   2s, and we term them as ``obscured NLS1'' galaxies to better
   reflect their nature.  The existence of obscured NLS1s is
   consistent with the standard unification of Seyfert galaxies.
   However, the presence of NLS1 nuclei in both type 1 and 2 Seyferts
   requires a new parameter, the relative accretion rate, to be
   incorporated in the standard unification scheme based on the
   viewing angle.
 
   We thank an anonymous referee for valuable comments. We thank N. J.
   Schurch for the discussions on NGC~7582, Tahir Yaqoob on the
   classification of NGC~7314, and A. R. Rao for going through the
   manuscript. GCD acknowledges the support of NASA grant through the
   award NNG04GN69G.


\begin{thebibliography}{}

   \bibitem[Antonucci(1993)]{1993ARA&A..31..473A} Antonucci, R.\ 1993,
     \araa, 31, 473
   \bibitem[Aretxaga et al.(1999)]{1999ApJ...519L.123A} Aretxaga, I.,
     Joguet, B., Kunth, D., Melnick, J., \& Terlevich, R.~J.\ 1999,
     \apjl, 519, L123
   \bibitem[Bianchi et al.(2003)]{2003A&A...402..141B} Bianchi, S.,
     Balestra, I., Matt, G., Guainazzi, M., \& Perola, G.~C.\ 2003,
     \aap, 402, 141
   \bibitem[Boller et al.(1996)]{1996A&A...305...53B} Boller, T.,
     Brandt, W.~N., \& Fink, H.\ 1996, \aap, 305, 53
   \bibitem[Condon et al.(1998)]{1998AJ....116.2682C} Condon, J.~J.,
     Yin, Q.~F., Thuan, T.~X., \& Boller, T.\ 1998, \aj, 116, 2682
   \bibitem[Brandt et al.(1997)]{1997MNRAS.285L..25B} Brandt, W.~N.,
     Mathur, S., \& Elvis, M.\ 1997, \mnras, 285, L25
   \bibitem[de Robertis \& Osterbrock(1986)]{1986ApJ...301..727D} de
     Robertis, M.~M., \& Osterbrock, D.~E.\ 1986, \apj, 301, 727
   \bibitem[Fabian(1979)]{1979RSPSA.366..449F} Fabian, A.~C.\ 1979,
     Royal Society of London Proceedings Series A, 366, 449
   \bibitem[Hughes et al.(2003)]{2003AJ....126..742H} Hughes, M.~A.,
     et al.\ 2003, \aj, 126, 742
   \bibitem[Joguet et al.(2001)]{2001A&A...380...19J} Joguet, B.,
     Kunth, D., Melnick, J., Terlevich, R., \& Terlevich, E.\ 2001,
     \aap, 380, 19
   \bibitem[Leighly(1999)]{1999ApJS..125..297L} Leighly, K.~M.\ 1999a,
     \apjs, 125, 297
   \bibitem[Leighly(1999)]{1999ApJS..125..317L} Leighly, K.~M.\ 1999b,
     \apjs, 125, 317
   \bibitem[Magdziarz \& Zdziarski(1995)]{1995MNRAS.273..837M}
     Magdziarz, P., \& Zdziarski, A.~A.\ 1995, \mnras, 273, 837
   \bibitem[Matt et al.(2001)]{2001A&A...377L..31M} Matt, G.,
     Guainazzi, M., Perola, G.~C., Fiore, F., Nicastro, F., Cappi, M.,
     \& Piro, L.\ 2001, \aap, 377, L31
   \bibitem[Morris \& Ward(1988)]{1988MNRAS.230..639M} Morris, S.~L.,
     \& Ward, M.~J.\ 1988, \mnras, 230, 639
   \bibitem[Nagar et al.(2002)]{Nagaretal02} Nagar, N.~M., Oliva, E.,
     Marconi, A., \& Maiolino, R.\ 2002, \aap, 391, L21
   \bibitem[Pounds et al.(1995)]{1995MNRAS.277L...5P} Pounds, K.~A.,
     Done, C., \& Osborne, J.~P.\ 1995, \mnras, 277, L5
   \bibitem[Risaliti(2002)]{2002A&A...386..379R} Risaliti, G.\ 2002,
     \aap, 386, 379
   \bibitem[Shuder(1980)]{1980ApJ...240...32S} Shuder, J.~M.\ 1980,
     \apj, 240, 32
   \bibitem[Sosa-Brito et al.(2001)]{2001ApJS..136...61S} Sosa-Brito,
     R.~M., Tacconi-Garman, L.~E., Lehnert, M.~D., \& Gallimore,
     J.~F.\ 2001, \apjs, 136, 61
   \bibitem[Storchi-Bergmann \& Bonatto(1991)]{1991MNRAS.250..138S}
     Storchi-Bergmann, T., \& Bonatto, C.~J.\ 1991, \mnras, 250, 138
   \bibitem[Turner(1987)]{1987MNRAS.226P...9T} Turner, T.~J.\ 1987,
     \mnras, 226, 9P
   \bibitem[Turner et al.(2000)]{2000ApJ...531..245T} Turner, T.~J.,
     Perola, G.~C., Fiore, F., Matt, G., George, I.~M., Piro, L., \&
     Bassani, L.\ 2000, \apj, 531, 245
   \bibitem[Veron et al.(1980)]{1980A&A....87..245V} Veron, P.,
     Lindblad, P.~O., Zuiderwijk, E.~J., Veron, M.~P., \& Adam, G.\
     1980, \aap, 87, 245
   \bibitem[Whittle(1985)]{1985MNRAS.216..817W} Whittle, M.\ 1985,
     \mnras, 216, 817
   \bibitem[Yaqoob et al.(1996)]{1996ApJ...470L..27Y} Yaqoob, T.,
     Serlemitsos, P.~J., Turner, T.~J., George, I.~M., \& Nandra, K.\
     1996, \apjl, 470, L27
   \bibitem[Yaqoob et al.(2003)]{2003ApJ...596...85Y} Yaqoob, T.,
     George, I.~M., Kallman, T.~R., Padmanabhan, U., Weaver, K.~A., \&
     Turner, T.~J.\ 2003, \apj, 596, 85
   \bibitem[Winkler(1992)]{1992MNRAS.257..677W} Winkler, H.\ 1992,
     \mnras, 257, 677
   \end{thebibliography}
\end{document}